\documentclass[11pt,letterpaper,twocolumn]{article} %% two column, final layout

\usepackage{ol2,mathptmx}
\usepackage{mathptmx}
\usepackage[draft,implicit=false]{hyperref}
\usepackage{amsmath,amssymb,graphicx}
\usepackage{float}

\newcommand{\be}{\begin{equation}}
\newcommand{\ee}{\end{equation}}
\newcommand{\ben}{\begin{equation*}}
\newcommand{\een}{\end{equation*}}
\newcommand{\bea}{\begin{eqnarray}}
\newcommand{\eea}{\end{eqnarray}}

\newcommand{\pd}{\partial}

\newcommand{\mum}{~\mu\text{m}}

\newcommand{\bbu}{~\text{ps}^2/\text{km}}
\newcommand{\bbbu}{~\text{ps}^3/\text{km}}

\begin{document}

\twocolumn[

\title{Femtosecond Pulse Trains through Dual-Pumping of Optical Fibers: \\ Role of Third-Order Dispersion}

\author{Aku Antikainen}
\address{
aku.antikainen@rochester.edu \\
The Institute of Optics, University of Rochester, Rochester, New York 14627, USA
}

\author{Govind P. Agrawal}
\address{
The Institute of Optics, University of Rochester, Rochester, New York 14627, USA \\
Laboratory for Laser Energetics, 250 East River Rd, Rochester, NY 14623, USA
}

\begin{abstract}
\textbf{Abstract:} Generation of high-repetition-rate, femtosecond, soliton pulse trains through dual-wavelength pumping of a dispersion-decreasing fiber is studied numerically. The achievable shortest pulse width is found to be limited by third-order dispersion that has a significant effect on the pulse-compression dynamics. The output wavelength is red shifted because of intrapulse Raman scattering and depends heavily on third-order dispersion, whose positive values lead to the most red shifted solitons ($>25$\% of the input pump center wavelength). The proposed scheme allows the generation of ultrashort pulse trains at tunable high repetition rates with a wide range of output wavelengths and pulse durations through dispersion engineering. The resulting frequency combs extend over a wide bandwidth with a tunable spacing between the comb lines.\newline \newline
\end{abstract}

%\tiny{OCIS codes: (320.5520) Pulse compression; (320.7140) Ultrafast processes in fibers; (190.4370) Nonlinear optics, fibers; (190.5530) Pulse propagation and temporal solitons; (320.2250) Femtosecond phenomena}
%\newline

                          % FORMAT THE TERMS CORRECTLY FOR EACH JOURNAL

%\thispagestyle{fancy}
]

\section{Introduction}

Soliton compression in active fibers \cite{Cao-2003a, Rosanov-2008} and dispersion-decreasing fibers (DDFs) \cite{Chernikov-1991a, Chernikov-1993a, Pelusi-1997, Fatemi-2002, Lee-2004, Hu-2006} is a well-understood effect \cite{Grimshaw-1979, Kuehl-1988}. It is known that the characteristics of compressed solitons can be controlled by tailoring fiber dispersion. In the extreme, this process can lead to the generation of intense solitons with durations of only a few optical cycles \cite{Tamura-2001}. Moreover, due to the tendency of pulses of suitable peak powers to re-adjust inside the fiber to form solitons, soliton compression does not necessarily have to be seeded by solitonic pulses. Indeed, pulses of Gaussian or other shapes can be used as input to the DDF\@. When the input is in the form of a continuous wave (CW) or a long pulse, modulation instability (MI) can lead to the spontaneous generation of fundamental solitons that can then be compressed.

An effective way to enhance soliton formation from a CW input is to seed the MI process through amplitude modulation \cite{Hasegawa-1984}. Induced MI in a fiber with constant (anomalous) dispersion leads to compression of low-amplitude temporal modulations, eventually resulting in a train of solitons. The formation of solitons, as well as their further compression, manifests as spectral broadening in the frequency domain that is useful for supercontinuum generation \cite{Book:Agrawal, Mori-1997, Okuno-1998}. However, pulse compression is not the only spectral broadening mechanism in the dual- or multi-pumping case \cite{Trillo-1994}. Amplitude modulation can lead to significantly broader spectra in the normal dispersion regime as well due to enhanced self-phase modulation and optical wave-breaking \cite{Champert-2004, Finot-2008, Antikainen-2015}, and temporal reflections of dispersive waves off nonlinear waves can extend the spectrum to the blue side \cite{Demircan-2014, Antikainen-2016, Antikainen-2017}. Modulation can be done either through direct amplitude-modulation of a single CW laser \cite{Lee-2004} or through dual-wavelength pumping using two CW lasers \cite{Tai-1986, Chernikov-1992, Chernikov-1993b, Pitois-2002}.

The frequency of direct amplitude modulation is limited by electronics since amplitude modulators cannot operate effectively beyond $40$~GHz. Dual-wavelength pumping suffers from no such constraints and is only limited by the availability of CW lasers of suitable wavelengths. Moreover, modulation depths depend only on the relative powers of the two pumps. Therefore, ultrahigh repetition rate ($800$~GHz in this study) pulse trains can be generated through dual-pumping of an optical fiber. Such pulse trains have a variety of applications ranging from controlling the motion of molecules \cite{Weiner-1990} to generating plasma waves \cite{Umstadter-1994} and terahertz radiation \cite{Liu-1996}. The objective of this paper is two-fold. First, we show that the lower limit for pulse duration in dual-pump soliton train generation is determined by higher-order dispersion; in the absence of third-order dispersion (TOD) the pulses could be compressed down to a few cycles in duration in a suitable DDF\@. Second, we show that the sign and magnitude of the TOD plays a crucial role in determining the wavelength of the generated train of solitons and that the output pulses can be red shifted by more than $25$\% from the initial pump center wavelength. Sub-$100$~fs soliton trains can thus be generated at a wide range of wavelengths by properly engineering the dispersion profile of the fiber. Our findings should help in designing fiber-based, high-repetition-rate, femtosecond-pulse sources and wide-band, optical frequency combs with a tunable spacing between its comb lines.

\section{Pulse Propagation Model}

To simulate propagation of electromagnetic waves in optical fibers we use the generalized nonlinear Schr{\"o}dinger equation (GNLSE) \cite{Book:Agrawal,Dudley-2006}. In a reference frame moving at the envelope group velocity, the equation for the electric-field envelope $A(z,t)$ can be written as
\begin{align}
	&\frac{\pd A}{\pd z} + \frac{\alpha}{2} A -  \sum_{n \geqslant 2} \frac{i^{n+1}}{n!} \beta_n \frac{\pd^n A}{\pd T^n}
	= i \gamma \left(1 + i \tau_\text{shock} \frac{\pd}{\pd T} \right) \cdot  \nonumber \\
	&\left( A(z,T) \int_{-\infty}^{\infty} R(T') \left| A(z,T-T')\right|^2 dT' \right) ,
	\label{eq:GNLSE}
\end{align}
where $T$ is the retarded time given by $T = t - z/v_g$ and $v_g$ is the group velocity at the input central wavelength. The left side of Eq.\ (\ref{eq:GNLSE}) includes linear effects with $\alpha$ corresponding to losses and the $\beta$'s being the different-order dispersion coefficients of the fiber at the central input frequency. The right side models the nonlinearities through the response function $R(T)$ that includes the Kerr contribution that is assumed instantaneous \cite{Book:Agrawal} and the delayed Raman contribution that is modeled through the experimental Raman-gain profile of silica fibers \cite{Stolen-1989}. Note that, contrary to a common misconception, the GNLSE for fibers does not assume a slowly-varying envelope for the electric field in the time domain and is in fact valid down to the few- and even single-cycle regime, as long as the wavelengths in question are far from material resonances so that the \emph{slowly-evolving-wave approximation} $|\pd_z E| \ll \beta_0 |E|$ is satisfied, as shown by Brabec and Krausz \cite{Brabec-1997}. The validity of the model down to the few-cycle regime requires the inclusion of the shock term characterized by the time scale $\tau_\text{shock}$ \cite{Dudley-2006}.

We solve Eq.\ (1) numerically with an input corresponding to launching two CW pumps simultaneously with a frequency difference $\Delta\omega_0$:
\begin{equation}
	A(0,T) = \sqrt{P_1}e^{i(\omega_0+\Delta\omega_0/2)t} + 
    \sqrt{P_2}e^{i(\omega_0-\Delta\omega_0/2)t}
\end{equation}
where $\omega_0$ is the central frequency and $P_1$ and $P_2$ are the input powers of the two pumps. In our simulations, the group-velocity dispersion (GVD) parameter $\beta_2$ increases linearly from its initial negative (anomalous) value of $-10~\text{ps}^2/\text{km}$ over the entire fiber length that varies in the range $100$ - $200$~m. The final values of $\beta_2$ range from $-10$ to $+10~\text{ps}^2/\text{km}$, the former corresponding to a constant-dispersion fiber. The TOD parameter $\beta_3$ is kept constant for each fiber with values ranging from $-0.1$ to $0.1~\text{ps}^2/\text{km}$. It should be noted that usually DDFs are manufactured by tapering the fibers, in which case the taper can induce significant losses and thus impose limitations on pulse compression \cite{Hu-2006}. However, dispersion can also be modified through doping, allowing for the losses to be curbed. The nonlinear parameter in our simulations is $\gamma=0.0916~1/(\text{Wm})$, and the pump powers are taken to be equal with $P_1=P_2=1$~W\@. The center frequency $\omega_0$ corresponds to a wavelength of 1060~nm. The shock time $\tau_\text{shock}$ is taken to be $1/\omega_0$. Throughout this paper the frequency separation is $\Delta\omega/(2 \pi) = 800$~GHz (3~nm) but we have also verified that frequency separations of $600$~GHz and $1000$~GHz yielded qualitatively similar results.

\section{Theoretical Limits on Pulse Width}

Fundamental solitons are solutions of the GNLSE in the absence of third- and higher-order dispersion, optical shock effects, and delayed nonlinearities. When these effects are present, they manifest as perturbations to an ideal soliton. Third-order dispersion (TOD) governed by $\beta_3$ introduces spectral and temporal asymmetry and forces the soliton to shed radiation in the form of a dispersive wave. Shock effects produce self-steepening, again making the soliton asymmetric. Delayed nonlinearities lead to the well-known phenomenon of soliton self-frequency shift (SSFS) through intrapulse Raman scattering, causing the soliton to red shift in the spectral domain. Nevertheless, the robust solitonic nature of the pulse remains. Solitons are robust to the extent that any pulse of suitable shape and energy in the anomalous dispersion regime of a nonlinear fiber will reshape itself to become one \cite{Book:Agrawal}. In the context of a beating dual-pump signal the sinusoidal oscillations at the beat frequency become compressed and evolve to form solitons if their duration and energy roughly matches those of a fundamental soliton. If the fiber is long enough, the beating intensity pattern eventually evolves to become a train of equidistant solitons. By changing the dispersion along the length of a DDF, the solitons can be compressed further in the temporal domain \cite{Chernikov-1991a}. 

The dispersion parameters $\beta_n$ in Eq.~(\ref{eq:GNLSE}) can be easily tailored through proper design of the refractive index profile, which in the case of photonic crystal fibers means appropriately choosing the size and spacing of the air holes surrounding the core. The only limitations regarding the structure of silica-based photonic crystal fibers are associated with manufacturing precision. In general, different photonic crystal fiber structures would also lead to different nonlinear coefficients for the fibers. However, since it is the relative strength of dispersion and nonlinearity that determines the propagation of light, we assume here that the nonlinear parameter is constant while $\beta_2$ changes linearly along the fiber. We also note that ultra-flat highly anomalous dispersion profiles can be achieved over a wide wavelength range with novel designs \cite{Islam-2012}. The TOD and other higher-order dispersion terms play a relatively minor role for such fibers.

\begin{figure}[t]
  \begin{center}
    \includegraphics[width=\columnwidth, trim = 0mm 0mm 0mm 0mm, clip]{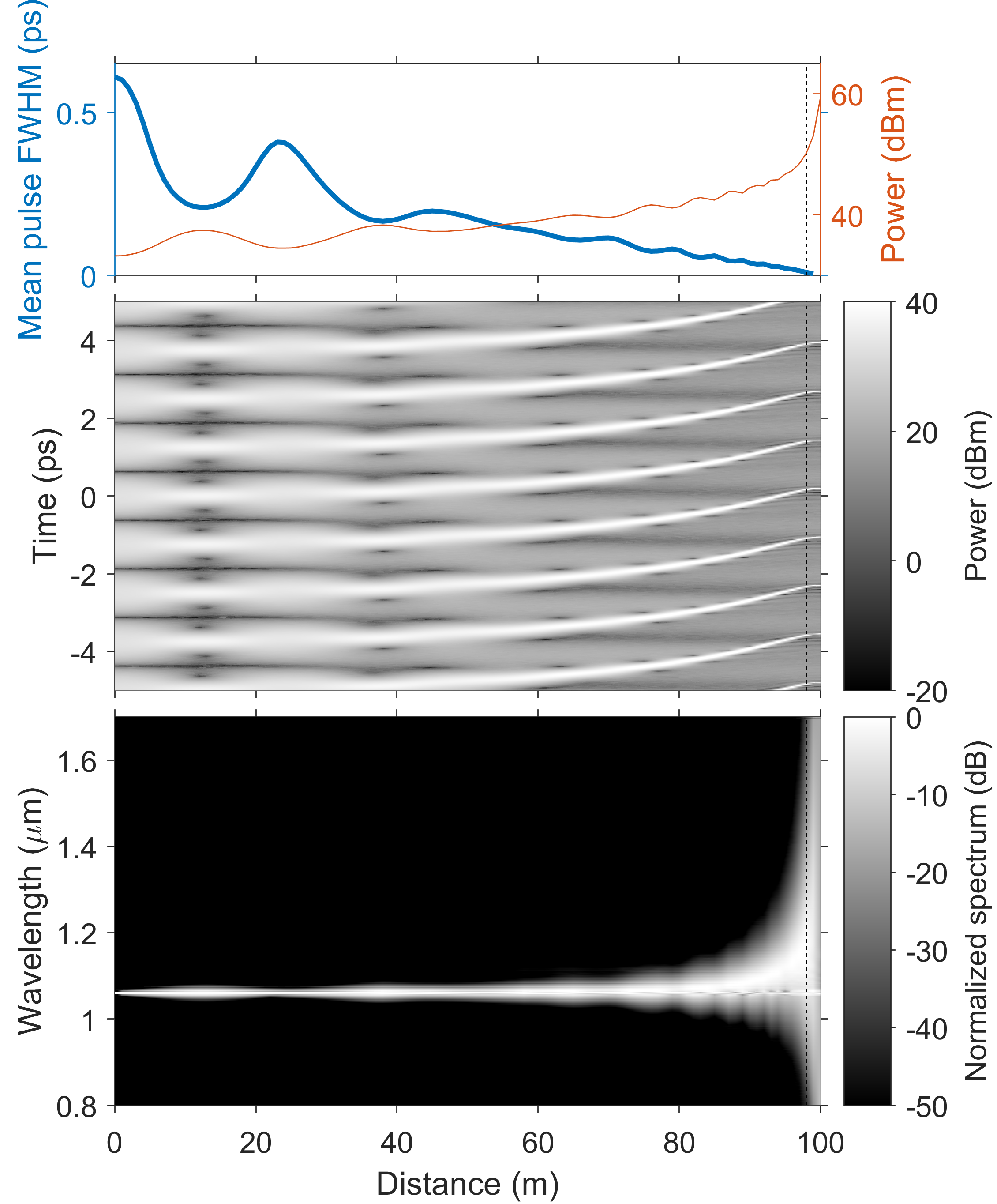}
		\caption{The temporal (middle) and spectral (bottom) evolution of a dual-pump signal over 100 meters of a DDF with $\beta_2$ increasing from $-10~\text{ps}^2/\text{km}$ to $0~\text{ps}^2/\text{km}$. The gray intensity scales are logarithmic. The top two traces show the duration (thick blue) and peak power (thin red) of the forming pulses as a function of distance. The vertical black dashed lines indicate the distance at which the soliton width has been reduced to three optical cycles.}
		\label{fig1}
  \end{center}
\end{figure}

To understand the dynamics of a dual-wavelength signal inside a DDF, we first neglect the TOD and other higher-order dispersion terms so that the effects of a longitudinally varying $\beta_2$ can be identified clearly. Figure \ref{fig1} shows the evolution of a dual-pump signal when $\beta_2$ increases linearly from $-10$ to $0~\text{ps}^2/\text{km}$ over 100~m. The power of both pumps is $1$~W. The two traces on top show changes in the pulse width and peak powers over the 100~m length of the fiber. The initial sinusoidal pattern gradually reshapes into a train of solitons whose width decreases and peak power increases continuously until the numerical model itself breaks down. The spectrum of the resulting pulse train is in the form of a frequency comb whose bandwidth is inversely related to the width of solitons and exceeds 100~THz. 

The compression dynamics in Figure \ref{fig1} have interesting features. The initial sinusoidal pattern with a period of 1.25~ps evolves into a pulse train within the first 10~m such that individual pulses are about 200~fs wide (full width at half maximum or FWHM). These soliton-like pulses then broaden with further propagation before being compressed a second time. This process repeats a few times but the pulse duration keeps a downward trend while exhibiting transient oscillations. The evolution of the solitons is affected by two mechanisms. First, varying fiber dispersion forces them to compress. Second, at the  same time, their speed is reduced as their spectrum red shifts because of SSFS (leading to bending of the trajectories in Figure \ref{fig1}). The individual solitons grow in intensity because of the increasing $\beta_2$, but also because they feed off the darker regions (energy in the low-intensity parts) when they shift in time and overlap temporally with them. This mode of energy transfer to the solitons is evident in Fig.\ \ref{fig1}, where the regions between the neighboring solitons become darker as the solitons slow down and pass through these regions. This energy transfer perturbs the solitons, causing their widths and peak powers to oscillate around their respective trends (decreasing duration, increasing peak power). 
%The oscillations have a smaller amplitude between $5$~m and $6.5$~m when the solitons are in the dark regions with limited interactions with the pump remains but increase again when the solitons enter the next brighter background after $6.5$ meters of propagation.

The GNLSE model given in Eq. (\ref{eq:GNLSE}) accurately describes pulse propagation down to the single-cycle regime \cite{Brabec-1997, Dudley-2006}, and in this study the three-cycle point is used as the cutoff for the validity of the GNLSE model. The distance at which the solitons in Fig.\ \ref{fig1} have compressed to three optical cycles in duration (about 10~fs) is approximately $98$ meters, and this has been indicated by the vertical dashed lines in Fig.\ \ref{fig1}. The important takeaway from Fig.\ \ref{fig1} is that the initial beating intensity pattern with a period of 1.25~ps (corresponding to $800$~GHz) could ideally be reshaped into a train of solitons that are only three optical cycles long. The input FWHM of the cosine-shaped pulses is $625$~fs, implying that the compression factor is larger than 50.

The simulation shown in Fig.\ \ref{fig1} includes all the relevant effects that would be present in reality, with the exception of higher-order dispersion and losses. The fiber was assumed to be lossless and to have perfectly flat dispersion (constant $\beta_2$) over all wavelengths at any given point of the fiber. Therefore, Fig.\ \ref{fig1} represents the best case scenario in terms of how short the solitons can become: Under ideal conditions pulse durations of three optical cycles or less could be achieved. Several different effects might prevent such drastic compression in practice, but the extent of compression is not limited by GVD, intrapulse Raman scattering, or optical shock effects. Losses would cause the peak power $P_0$ of the forming solitons to be smaller, which in turn would lead to larger soliton durations $T_0$ such that the soliton condition of $\gamma P_0 T_0^2 / \beta_2(z) = 1$ continues to be satisfied. However, at the end of the fiber $\beta_2 = 0$, and the soliton condition can only be satisfied for infinitely narrow solitons no matter what the peak power might be. Compensation for losses through decreasing dispersion (increasing $\beta_2$) to keep the soliton duration unchanged upon propagation in lossy fibers has been demonstrated in the past \cite{Richardson-1995}. Decreasing dispersion even faster than in Fig.~1 would be required to compensate for any possible fiber losses. However, TOD could be expected to change the compression dynamics more drastically than losses because it affects solitons in at least three different ways: It leads to dispersive-wave emission, it asymmetrically distorts the shape of a soliton, and it makes $\beta_2$ frequency-dependent.

\section{Effects of Third Order Dispersion}

The first thing to note is that the sign of TOD plays an important role in the evolution of a short solitons undergoing intrapulse Raman scattering. The SSFS causes the soliton spectrum to red shift, and it is the sign of $\beta_3$ that then determines whether the soliton will experience a larger or smaller $\beta_2$ as a consequence. Since soliton compression is based on increasing $\beta_2$ from an initially negative value through dispersion engineering, any TOD-induced change to $\beta_2$ will affect the compression of solitons. The presence of TOD also introduces a spectral region or normal dispersion in which solitons cannot exist but also guarantees the existence of a spectral region of anomalous dispersion even when $\beta_2 > 0$ at the pump frequency. The signs of $\beta_2$ and $\beta_3$ determine whether the normal dispersion regime is on the red or the blue side of the soliton. The frequency at which GVP changes sign is given by $\omega_\text{ZDW} = \omega_0 - \beta_2/\beta_3$ where $\beta_2$ and $\beta_3$ are evaluated at the central frequency $\omega_0$. The wavelength corresponding to $\omega_\text{ZDW}$ is the zero-dispersion wavelength (ZDW). When $\beta_2$ is a linear function of distance $z$ we have
\begin{equation}
	\beta_2(\omega_0) = \beta_2^\text{in} + (\beta_2^\text{out} 
    - \beta_2^\text{in})\frac{z}{L}
\end{equation}
where $L$ is the length of the fiber and $\beta_2^\text{in}$ and $\beta_2^\text{out}$ are the output values of $\beta_2$ at $\omega_0$. Consequently the ZDW becomes a function of $z$ through
\begin{equation}
	\omega_\text{ZDW} = \omega_0 -\frac{\beta_2^\text{in}}{\beta_3} 
    - (\beta_2^\text{out} - \beta_2^\text{in})\frac{z}{\beta_3L}.
	\label{eqzdw}
\end{equation}

\begin{figure}[ht]
  \begin{center}
    \includegraphics[width=\columnwidth, trim = 0mm 0mm 0mm 0mm, clip]{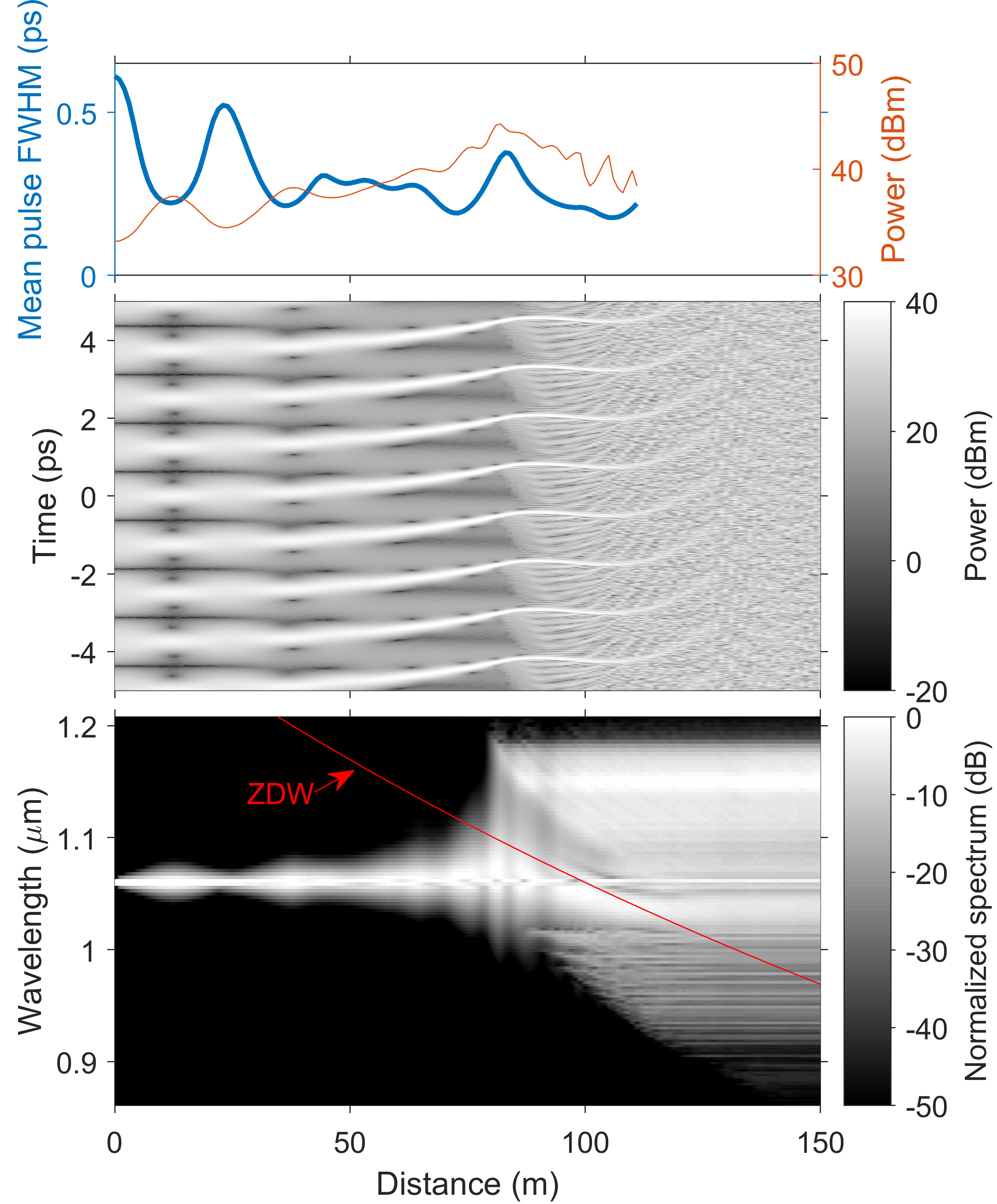}
		\caption{The evolution of a $800$~GHz dual-pump signal in a fiber in which $\beta_2$ grows from $-10 \bbu$ to $5 \bbu$ along its $150$~m length. Third-order dispersion is $\beta_3 = -0.03 \bbu$.}
		\label{fige1}
  \end{center}
\end{figure}

To illustrate the effects of TOD in a DDF, Fig.\ \ref{fige1} shows the evolution in a fiber where $\beta_2$ changes from $-10 \bbu$ to $5 \bbu$ over $150$ meters and where $\beta_3 = -0.03 \bbbu$. Note that the rate of change of $\beta_2$ with $z$ is the same as for the fiber in Fig.\ \ref{fig1} and the ZWD coincides with the pump center wavelength at exactly $100$ meters just like in Fig.\ \ref{fig1}.

The evolution of the dual-pump shown in Fig.\ \ref{fige1} differs from that of Fig.\ \ref{fig1}. The most noticeable difference between the two cases is that the pulses do not become infinitely narrow when $\beta_3 \neq 0$ and the minimum pulse duration in Fig.\ \ref{fige1} is approximately $180$~fs. The formation of few-cycle pulses would require a very broad pulse spectrum and since the pulses are solitons this spectrum would have to lie in the anomalous dispersion regime. When $\beta_3 \neq 0$ and the ZDW approaches the soliton spectrum, the tail of the pulse spectrum will eventually end up in the normal dispersion regime thus limiting the spectral extent and consequently the pulse duration of the solitons. The first effects can be observed after $80$~m of propagation when the ZDW starts to touch the tail of the soliton spectrum and power is transferred from the solitons to a dispersive wave on the red side of the ZDW. The moving ZDW gradually puts more and more energy to the normal dispersion regime and the soliton peak powers start to decrease. The ZDW crosses the center of the soliton spectrum around $110$~m and after this the solitons cease to exist and disperse into a chaotic-looking yet nearly-periodic pattern of interfering waves in the normal dispersion regime. After this point it is no longer meaningful to talk about soliton peak powers or durations or consider the intensity profile a train of pulses.

\begin{figure}[ht]
  \begin{center}
    \includegraphics[width=\columnwidth, trim = 0mm 0mm 0mm 0mm, clip]{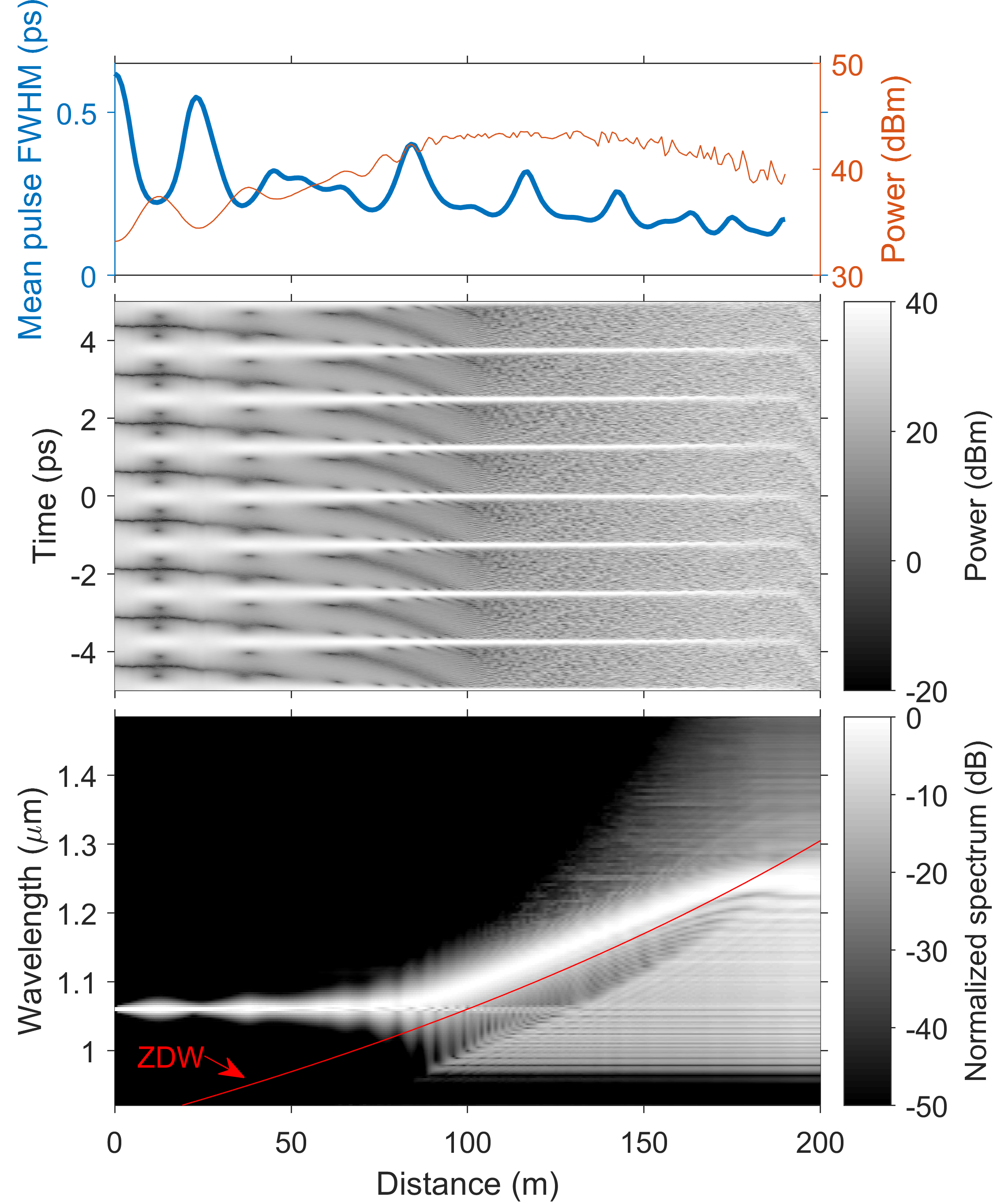}
		\caption{The evolution of a $800$~GHz dual-pump signal in a fiber in which $\beta_2$ grows from $-10 \bbu$ to $10 \bbu$ along its $200$~m length. Third-order dispersion is $\beta_3 = 0.03 \bbbu$. Unlike in Figs.\ \ref{fig1} and \ref{fige1}, the temporal frame of reference is now with respect to the solitons, as their trajectories would look heavily curved in the pump frame of reference.}
		\label{fige2}
  \end{center}
\end{figure}

In the example shown in Fig.\ \ref{fige1} the frequency slope of $\beta_2$ was negative ($\beta_3 < 0$) and hence the normal dispersion regime was on the red side of the pump. Solitons have a tendency to try to stay away from the ZDW and remain in the anomalous regime, which can be seen in the spectrum of Fig.\ \ref{fige1} where the spectral trajectory of the soliton bends slightly downwards between $90$~m and $110$~m and the solitons blue shift. The blue shift is always accompanied by significant transfer of energy to the red side of the ZDW to conserve total energy. Normally solitons, especially short ones, have a tendency to red shift upon propagation because of intrapulse Raman scattering. This raises the question whether having the ZDW approach the soliton spectrum from the blue side instead would help the solitons remain in the anomalous regime for longer distances. Figure~\ref{fige2} shows the evolution of a $800$~GHz dual-pump in a fiber with $\beta_3 = 0.03 \bbbu$. Other than the fiber length and the TOD, the fiber is similar to the ones in Figs.\ \ref{fig1} and \ref{fige1} and again the ZDW is at the pump center at $100$~m. Note that the temporal trace in Fig.~\ref{fige2} is now in the reference frame of the solitons instead of moving at the group velocity at the pump frequency.

The evolution of the soliton power and duration is similar to that of Fig.\ \ref{fige1} but the solitons last longer and the spectral evolution looks very different. The ZDW is now on the blue side of the solitons and the ZDW approaching the soliton spectrum greatly enhances the natural SSFS pushing the soliton spectrum all the way to $1.25\mum$ from the initial $1.06\mum$. Still, the moving ZDW eventually overtakes the soliton spectrum and in the end the pulses end up in the normal dispersion regime and disperse. The minimum soliton duration is $125$~fs around $185$~m.

\begin{figure}[ht]
  \begin{center}
    \includegraphics[width=\columnwidth, trim = 0mm 0mm 0mm 0mm, clip]{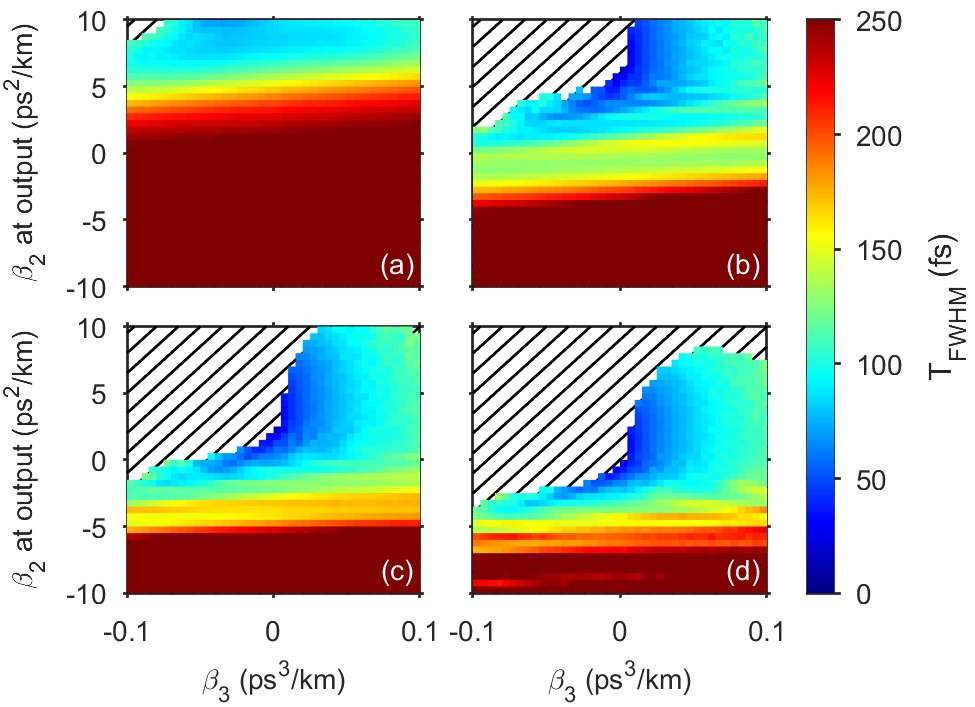}
		\caption{The mean duration (FWHM, color coded) of the forming solitons after a) $80$~m, b) $120$~m, c) $160$~m, and d) $200$~m of propagation as a function of $\beta_3$ and the final value of $\beta_2$. The initial value of $\beta_2$ at the input end of $200$-meter-long fiber is $-10~\text{ps}^2/\text{km}$. The striped areas in the upper left corners are regions where the pulses have lost their solitonic nature by virtue of having transferred energy to the normal dispersion regime.}
		\label{fig2}
  \end{center}
\end{figure}

To understand quantitatively the impact of $\beta_3$, we carried out a large number of numerical simulations for different DDF designs. Figure~\ref{fig2} shows the color-coded duration of solitons (range 0--250~fs) for $\beta_3$ values varying from $-0.1$ to $0.1~\text{ps}^3/\text{km}$ along the $x$ axis and different values of $\beta_2(L)$ at the end of a 200-m-long fiber with $\beta_2(0)=-10~ \text{ps}^2/ \text{km}$. In each case, $\beta_3$ is kept constant along the fiber. The four plots shows the soliton widths at distances of (a) 80, (b) 120, (c) 160, and (d) 200~m.

If the solitons forming from the beating input signal are able to keep up with the gradually changing GVD parameter $\beta_2$, then larger final values of $\beta_2$ leads to shorter solitons. The general trend in Fig.\ \ref{fig2} is that increasing the final value of $\beta_2$ makes the output pulses shorter, which means that solitons are mostly able to keep up with the longitudinally changing GVD, even when GVD becomes normal near the fiber end. This is also corroborated by Figs.\ \ref{fig1} where pulse duration has a downward linear trend approaching zero with decaying transient oscillations. The transient oscillations die out by the end of a $200$-meter-long fiber when the final value of $\beta_2$ is larger than $-5~\text{ps}^2/\text{km}$, as seen in Fig.\ \ref{fig2}. The temporal compression continues even after the oscillations disappear.

The effects of TOD are clearly visible in Fig.\ \ref{fig2}. Larger values of $|\beta_3|$ hinder pulse compression, whereas smaller values lead to shorter pulses at shorter distances. The explanation for this lies in how $\beta_3$ effects the $\beta_2$ that the soliton experiences and in the Raman effect that causes the soliton spectrum to red shift through SSFS with propagation. The TOD parameter is given by $\beta_3 = d\beta_2(\omega) / d\omega$ evaluated at the central frequency $\omega_0$. Negative values of $\beta_3$ thus mean that $\beta_2$ decreases with optical frequency and hence increases with wavelength. SSFS then causes the solitons to experience larger GVD compared to the initial pump center frequency. Negative values of $\beta_3$ together with SSFS imply that $\beta_2$ at the solitons' central frequency increases even faster than $\beta_2$ at the pump center frequency, thus causing the solitons to compress rapidly. The opposite occurs for positive values of $\beta_3$. As seen in Fig.\ \ref{fig1}, solitons could be compressed down the three optical cycles in the absence of TOD, but in practice pulse compression is limited by it. We note that fibers with $\beta_3 = 0$ can also be manufactured (so-called dispersion-flattened fibers); pulse compression would be limited by fourth-order dispersion. There is no way to make the group-velocity dispersion completely flat across the whole soliton spectrum and pulse compression will always be limited by higher-order dispersion.

\begin{figure}[tb!]
  \begin{center}
    \includegraphics[width=\columnwidth, trim = 0mm 0mm 0mm 0mm, clip]{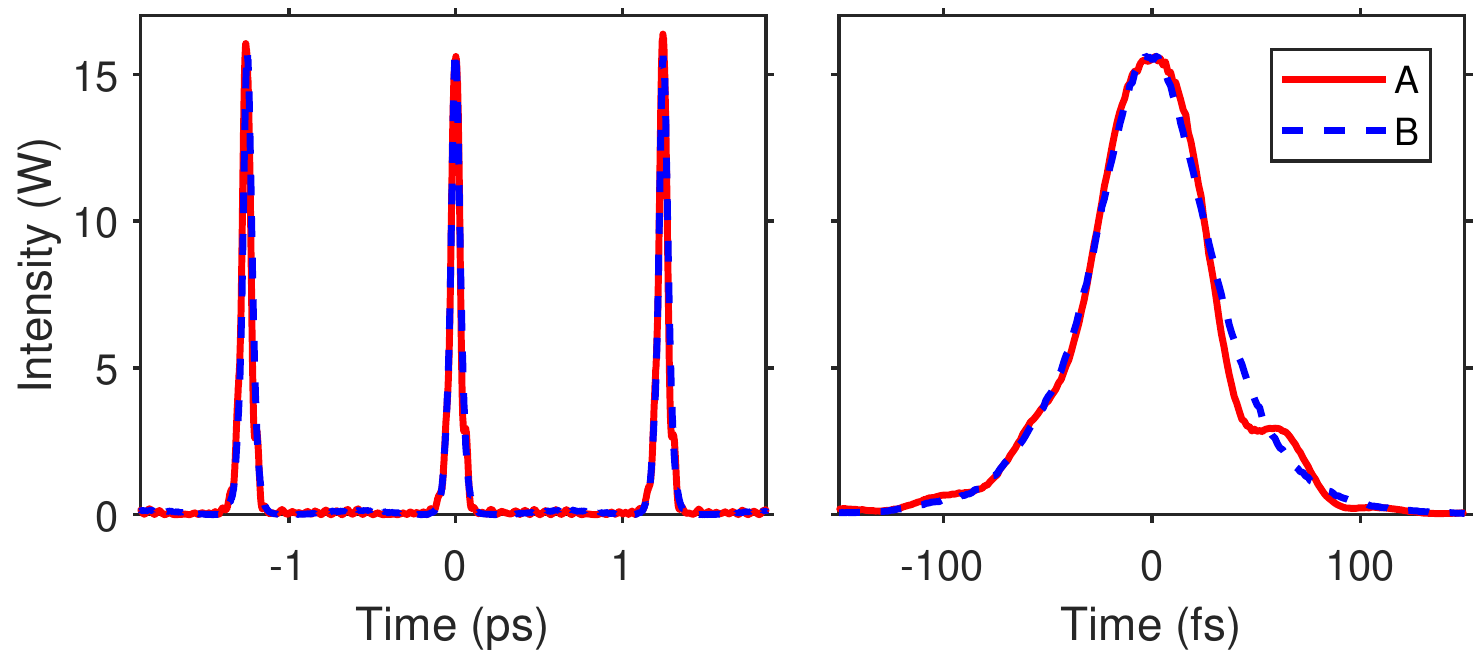}
		\caption{Comparison of pulse trains generated with the same dual-pump input in two different fibers. Fiber A is 100~m long and its GVD increases linearly from $-10~\text{ps}^2/ \text{km}$ to 0 over this length with $\beta_3 = 0.05~\text{ps}^3/\text{km}$. Fiber B is 97~m long but its GVD increases from $-10$ to $-2.725~\text{ps}^2/\text{km}$ with $\beta_3 = -0.05~\text{ps}^3/\text{km}$. The total input power is $2$~W and initial pump separation is $800$~GHz. The two traces on the right show the pulse around $T = 0$ showing how closely their shapes match.}
		\label{fig3}
  \end{center}
\end{figure}

It is evident from Fig.\ \ref{fig2} that soliton trains with pulse widths $< 100$~fs can be achieved with many different parameter combinations. Even a $100$-meter fiber can be long enough to produce such an ultrashort pulse train if $\beta_2$ of the DDF increases rapidly enough with distance [see Fig.\ \ref{fig2}(b)]. Both negative and positive values of $\beta_3$ work, and two different sets of fiber parameters can lead to very similar-looking pulse trains. Figure~\ref{fig3} shows portions of two pulse trains generated using two different fibers with the same input. Both fibers have the same GVD at the input end but their lengths and final values of $\beta_2$ are different. Their TOD parameters are equal in magnitude but opposite in sign. The solitons generated in each fiber are nearly identical: their energies and pulse durations are within $2$\% of one another. The only notable difference is that the pulses in the fiber with $\beta_3 > 0$ (Fiber A) exhibit a small bump near the trailing end. The differences between the pulse trains are subtle in the time domain but become quite evident in the spectral domain to which we turn in the next section.

\section{Output Frequency Comb and its Central Wavelength}

\begin{figure}[tb!]
  \begin{center}
    \includegraphics[width=\columnwidth, trim = 0mm 0mm 0mm 0mm, clip]{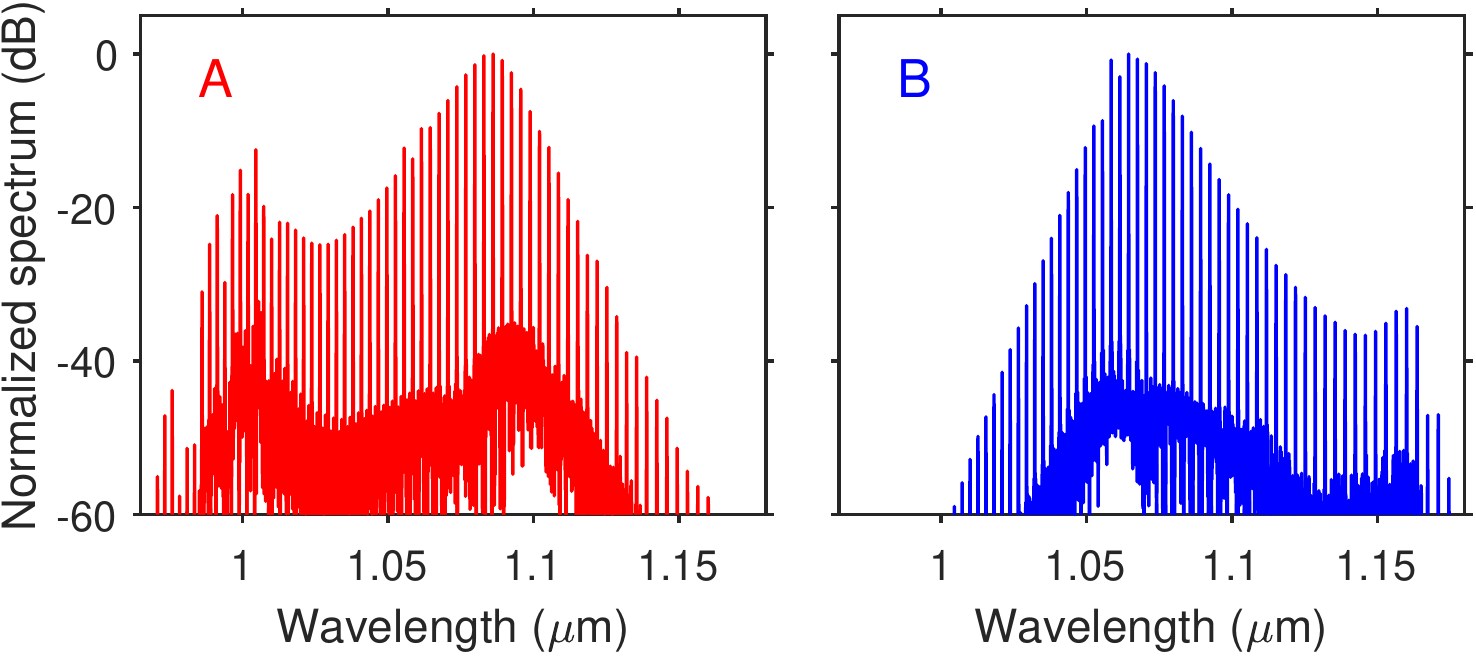}
		\caption{Spectra of the two pulse trains shown in Fig.\ \ref{fig3} at the output of fibers A and B.}
		\label{fig4}
  \end{center}
\end{figure}

The output spectrum of any periodic ultrashort pulse train generated through dual-pumping is in the form of a frequency comb whose comb lines are separated by the initial spacing between the frequencies of the two input pumps. Figure~\ref{fig4} shows the spectra corresponding to the two identical-looking pulse trains shown in Fig.\ \ref{fig3}. The spectra resemble mirror images of one another because of the opposite signs of the TOD parameter $\beta_3$. The soliton part of the spectrum (dominant peak) of fiber A is centered at 1086.1~nm, while that of fiber B is at 1067.5~nm, a difference of $18.6$~nm ($4.81$~THz). As a reminder, the input center wavelength of the two pumps is at 1060~nm.

The central frequency at each point in the fiber is determined by several processes. The first one is SSFS which causes the solitons to red shift. The second one is the tendency of solitons to stay away from the ZDW in the spectral domain \cite{Skryabin-2003}, and a moving ZDW can manifest as an effective push for the soliton spectrum. Depending on whether this push comes from the red side or the blue side, it can respectively hinder or enhance the red shift (See Figs.\ \ref{fige1} and \ref{fige2}, respectively). For $\beta_3 > 0$ we have $\omega_\text{ZDW} > \omega_0$, and $\omega_\text{ZDW}$ approaches $\omega_0$ from the blue side, enhancing the red shift and pushing the solitons further into the red. When $\beta_3 > 0$, $\omega_\text{ZDW}$ approaches $\omega_0$ from the red side and SSFS is thus hindered. This is the reson the spectrum out of fiber A in Fig.\ \ref{fig4} is more red shifted than that of fiber B.

\begin{figure}[ht]
  \begin{center}
    \includegraphics[width=\columnwidth, trim = 0mm 0mm 0mm 0mm, clip]{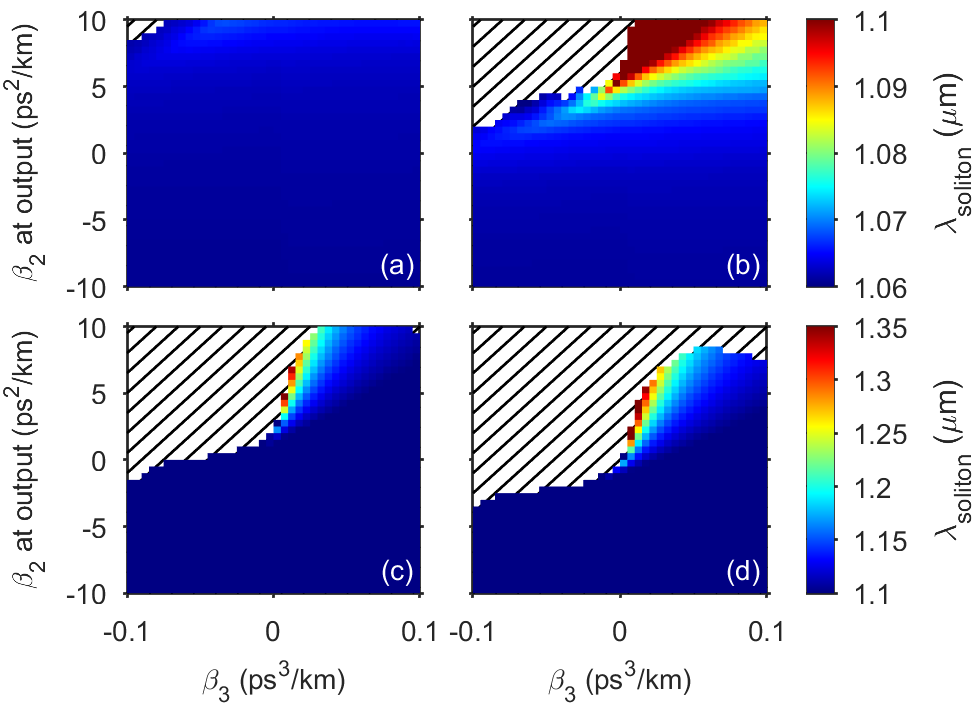}
		\caption{The central wavelengths $\lambda_\text{soliton}$ of the forming solitons for the parameters used in Fig.\ \ref{fig2} after a) $80$~m, b) $120$~m, c) $160$~m, and d) $200$~m of propagation. The striped regions indicate that the pulses have lost their solitonic nature and have dispersed. The upper colorbar is for the top row and the lower one for the bottom row; note the different scales.}
		\label{fig5}
  \end{center}
\end{figure}

If $\beta_2^\text{out} > 0$, $\omega_\text{ZDW}$ always surpasses $\omega_0$ no matter how fast or slow its rate of change. The rate of change is proportional to $1/\beta_3$ as seen in Eq.~(\ref{eqzdw}), which means that when $\beta_3$ is close to zero, $\omega_\text{ZDW}$ changes rapidly with distance $z$. Based on this argument, it seems likely that solitons could be pushed towards even longer wavelengths by making $\beta_3$ smaller while keeping it positive. Figure \ref{fig5} shows the central wavelength of the pulse trains generated through dual-pumping at distances of $80$~m, $120$~m, $160$~m, and $200$~m under conditions identical to those of Fig.\ \ref{fig2}. The initially forming solitons are wide at first and, as a result, red shifts of $<5$~nm occur up to a distance of 50~m. Much larger shifts occur at distances beyond $100$~m, especially for large values of $\beta_2^\text{out}$ for which adiabatic soliton compression kicks in and makes the solitons shorter thus enhancing their SSFS\@. The largest red shifts occur in the regime where $\beta_2^\text{out} > 0$ and $\beta_3$ is small but positive. The soliton central frequency can be red shifted by more than $25$\% to $1.35$~nm before $\omega_\text{ZDW}$ moves beyond the soliton central frequency and disperses the solitons.

\section{Conclusions}

It was numerically demonstrated that the technique of dual-wavelength pumping can be used to generate soliton pulse trains at ultrahigh-repetition rates (up to 1~THz or more) and that the solitons could be compressed temporally inside a dispersion-decreasing fiber down to the few-cycle regime (pulse widths as short as 10~fs at wavelengths near 1~$\mu$m). The repetition rate used in this study was $800$~GHz, but since it is set by the frequency separation of two CW pumps, it can be tuned over a wide range by choosing the input pump wavelengths suitably. It was further pointed out that the soliton compression is limited by higher-order dispersion with small values of the GVD slope $\beta_3 = d\beta_2/ d\omega$ leading to shortest pulses. It was also shown that third-order dispersion is crucial in determining the output wavelength of the pulses. We found that small positive values of the GVD slope lead to the largest red shifts and the longest output wavelengths. Sub-$100$~fs solitonic pulses with a wavelength anywhere between $1060$~nm and $1350$~nm could be achieved in our numerical simulations, making dual-wavelength pumped optical fibers a versatile platform for generating femtosecond pulses at high-repetition rates that have a variety of applications ranging from biomedical imaging to the manipulation of motion of individual molecules.

The spectral features of the generated pulse trains are also remarkable. Our results clearly show that the dual-pumping scheme is capable of generating frequency combs that extend over 50~THz and whose center frequency is tunable over $60$ THz in the vicinity of $1150$~nm. Moreover, the comb spacing in itself can be tuned over a wide range ($\sim$0.1 to $\sim$1~THz) by choosing the pump wavelengths suitably. As a final remark, the same technique should work for generating optical frequency combs from the visible to mid-infrared region using different fiber designs and materials.

%\section*{Full list of references}


\begin{thebibliography}{99}

\bibitem{Cao-2003a} W. Cao and P. K. A. Wai, ``Amplification and compression of ultrashort fundamental solitons in an erbium-doped nonlinear amplifying fiber loop mirror,'' Opt. Lett. {\bf 28}, 284--286 (2003)

\bibitem{Rosanov-2008} N. N. Rosanov, V. E. Semenov, and N. V. Vysotina, ``Few-cycle dissipative solitons in active nonlinear optical fibres,'' Quantum Electronics {\bf 38}(2), 2 (2008)

\bibitem{Chernikov-1991a} S. V. Chernikov and P. V. Mamyshev, ``Femtosecond soliton propagation in fibers with slowly decreasing dispersion,'' J. Opt. Soc. Am. B {\bf 8}, 1633--1641 (1991)

\bibitem{Chernikov-1993a} S. V. Chernikov, D. J. Richardson, D. N. Payne, and E. M. Dianov, ``Soliton pulse compression in dispersion-decreasing fiber,'' Opt. Lett. {\bf 18}, 476--478 (1993)

\bibitem{Pelusi-1997} M. D. Pelusi, H.-F. Liu, ``Higher order soliton pulse compression in dispersion-decreasing optical fibers,'' IEEE J. Quantum Electron. {\bf 33}(8), 1430--1439 (1997)

\bibitem{Fatemi-2002} F. K. Fatemi, ``Analysis of nonadiabatically compressed pulses from dispersion-decreasing fiber,'' Opt. Lett. {\bf 727}, 1637--1639 (2002)

\bibitem{Hu-2006} J. Hu, B. S. Marks, C. R. Menyuk, J. Kim, T.s F. Carruthers, B. M. Wright, T. F. Taunay, and E. J. Friebele, ``Pulse compression using a tapered microstructure optical fiber,'' Opt. Express {\bf 14}(9), 4026--4036 (2006)

\bibitem{Lee-2004} J. H. Lee, Y.-G. Han, S. B. Lee, T. Kogure, and D. J. Richardson, ``40 GHz adiabatic compression of a modulator based dual frequency beat signal using Raman amplification in dispersion decreasing fiber,'' Opt. Express {\bf 12}, 2187--2192 (2004)

\bibitem{Grimshaw-1979} R. Grimshaw, ``Slowly varying solitary waves. II. Nonlinear Schr{\"o}dinger equation,'' Proc. R. Soc. London Ser. A {\bf 368}(1734), 377 (1979)

\bibitem{Kuehl-1988} H. H. Kuehl, ``Solitons on an axially nonuniform optical fiber,'' J. Opt. Soc. Am. B {\bf 5}(3), 709--713 (1988)

\bibitem{Tamura-2001} K. R. Tamura and M. Nakazawa, ``54-fs, 10-GHz soliton generation from a polarization-maintaining dispersion-flattened dispersion-decreasing fiber pulse compressor,'' Opt. Lett. {\bf 26}, 762--764 (2001)

\bibitem{Hasegawa-1984} A. Hasegawa, ``Generation of a train of soliton pulses by induced modulational instability in optical fibers,'' Opt. Lett. {\bf 9}(7), 288--290 (1984)

\bibitem{Book:Agrawal} G. P. Agrawal, ``Nonlinear Fiber Optics,'' 5th ed. (Academic Press, 2013)

\bibitem{Mori-1997}	K. Mori, H. Takara, S. Kawanishi, M. Saruwatari, and T. Morioka, ``Flatly broadened supercontinuum spectrum generated in a dispersion decreasing fibre with convex dispersion profile,'' Electron. Lett. {\bf 33}(21), 1806--1808 (1997)

\bibitem{Okuno-1998}	T. Okuno, M. Onishi, and M. Nishimura, ``Generation of ultra-broad-band supercontinuum by dispersion-flattened and decreasing fiber,'' IEEE Photonics Technol. Lett. {\bf 10}(1), 72--74 (1998)

\bibitem{Trillo-1994} S. Trillo, S. Wabnitz, and T. A. B. Kennedy, ``Nonlinear dynamics of dual-frequency-pumped multiwave mixing in optical fibers,'' Phys. Rev. A {\bf 50}(2), 1732--1747

\bibitem{Champert-2004} P.-A. Champert, V. Couderc, P. Leproux, S. F{\'e}vrier, V. Tombelaine, L. Labont{\'e}, P. Roy, C. Froehly, and P. N{\'e}rin, ``White-light supercontinuum generation in normally dispersive optical fiber using original multi-wavelength pumping system,'' Optics Express {\bf 12}(19), 4366--4371 (2004)

\bibitem{Finot-2008} C. Finot, B. Kibler, L. Provost, and S. Wabnitz, ``Beneficial impact of wave-breaking for coherent continuum formation in normally dispersive nonlinear fibers '', J. Opt. Soc. Am. B {\bf 25}(11), 1938--1948 (2008)

\bibitem{Antikainen-2015} A. Antikainen and G. P. Agrawal, ``Dual-pump frequency comb generation in normally dispersive optical fibers,'' J. Opt. Soc. Am. B {\bf 32}, pp. 1705--1711 (2015)

\bibitem{Demircan-2014} A. Demircan, S. Amiranashvili, C. Br\'{e}e, U. Morgner, and G. Steinmeyer, ``Supercontinuum generation by multiple scatterings at a group velocity horizon,'' Opt. Express  {\bf 22}(4), 3866--3879 (2014)

\bibitem{Antikainen-2016} A. Antikainen, F. R. Arteaga Sierra, and Govind P. Agrawal, ``Supercontinuum Generation in Photonic Crystal Fibers with Longitudinally Varying Dispersion Using Dual-Wavelength Pumping,'' Frontiers in Optics, FTu1I.6 (2016)

\bibitem{Antikainen-2017} A. Antikainen, F. R. Arteaga-Sierra, and G. P. Agrawal, ``Temporal reflection as a spectral-broadening mechanism in dual-pumped dispersion-decreasing fibers and its connection to dispersive waves,'' Phys. Rev. A {\bf 95}(3), 033813 (2017)

\bibitem{Tai-1986} K. Tai, A. Tomita, J. L. Jewell, and A. Hasegawa, ``Generation of subpicosecond solitonlike optical pulses at 0.3 THz repetition rate by induced modulational instability,'' Appl. Phys. Lett {\bf 49}, 236--238 (1986)

\bibitem{Chernikov-1992} S. V. Chernikov, J. R. Taylor, P.V. Mamyshev, and E. M. Dianov, ``Generation of soliton pulse train in optical fibre using two cw singlemode diode lasers,'' Electron. Lett. {\bf 28}(10), 931--932 (1992)

\bibitem{Chernikov-1993b} S. V. Chernikov, E. M. Dianov, D. J. Richardson, R. I. Laming, and D. N. Payne, ``114 Gbit/s soliton train generation through Raman self-scattering of a dual frequency beat-signal in dispersion decreasing optical fibre,'' Appl. Phys. Lett {\bf 63}(3), 293--295 (1993)

\bibitem{Pitois-2002} S. Pitois, J. Fatome, and G. Millot, ``Generation of a 160-GHz transform-limited pedestal-free pulse train through multiwave mixing compression of a dual-frequency beat signal,'' Opt. Lett. {\bf 27}, 1729--1731 (2002)

\bibitem{Weiner-1990} A. M. Weiner, D. E. Leaird, G. P. Wiederrecht, K. A. Nelson, ``Femtosecond pulse sequences used for optical manipulation of molecular motion,'' Science {\bf 247}(4948), 1317--1319 (1990)

\bibitem{Umstadter-1994} D. Umstadter, E. Esarey, and J. Kim, ``Nonlinear Plasma Waves Resonantly Driven by Optimized Laser Pulse Trains,'' Phys. Rev. Lett. {\bf 72}(8), 1224--1227 (1994)

\bibitem{Liu-1996} Y. Liu, S.-G. Park, and A. M. Weiner, ``Enhancement of narrow-band terahertz radiation from photoconducting antennas by optical pulse shaping,'' Opt. Lett. {\bf 21}(21), 1762--1764 (1996)

\bibitem{Dudley-2006} J. M. Dudley, G. Genty, and S. Coen, ``Supercontinuum generation in photonic crystal fiber,'' Rev. Mod. Phys. {\bf 78}, 1135--1184 (2006)

\bibitem{Stolen-1989} R. H. Stolen, J. P. Gordon, W. J. Tomlinson, and H. A. Haus, ``Raman response function of silica-core fibers,'' J. Opt. Soc. Am. B {\bf 6}(6), 1159--1166 (1989)

\bibitem{Brabec-1997} T. Brabec and F. Krausz, ``Nonlinear Optical Pulse Propagation in the Single-Cycle Regime,'' Phys. Rev. Lett. {\bf 78}(17), 3282--3285 (1997)

\bibitem{Islam-2012} Md. A. Islam and M. S. Alam, ``Design Optimization of Equiangular Spiral Photonic Crystal Fiber for Large Negative Flat Dispersion and High Birefringence,'' J. Lightwave Technol. {\bf 30}(22), 3545--3551 (2012)

\bibitem{Richardson-1995} D. J. Richardson, R. P. Chamberlain, L. Dong, and D. N. Payne, ``High quality soliton loss-compensation in 38 km dispersion-decreasing fibre,'' Electron. Lett. {\bf 31}(19), pp. 1681--1682 (1995)

\bibitem{Skryabin-2003} D. V. Skryabin, F. Luan, J. C. Knight, P. St. J. Russell, ``Soliton Self-Frequency Shift Cancellation in Photonic Crystal Fibers,'' Science {\bf 301}(5640), pp. 1705--1708 (2003)


\end{thebibliography}
\end{document}